\documentclass[letter]{aa} 

\usepackage{amsmath}
\usepackage{graphicx}
\usepackage{graphics}
\usepackage{subfigure}
\usepackage{txfonts}
\usepackage{natbib}
\usepackage{xcolor}
\usepackage{tablefootnote}

\bibpunct{(}{)}{;}{a}{}{,}

\def\teff{\ifmmode T_{\rm eff} \else $T_{\mathrm{eff}}$\fi}

\def\ltsima{$\buildrel<\over\sim$}
\def\lsim{\lower.5ex\hbox{\ltsima}}

\newcommand{\hii}{H~{\sc ii}}
\newcommand{\ha}{\ifmmode {\rm H}\alpha \else H$\alpha$\fi}
\newcommand{\hb}{\ifmmode {\rm H}\beta \else H$\beta$\fi}
\newcommand{\lya}{\ifmmode {\rm Ly}\alpha \else Ly$\alpha$\fi}

\newcommand{\ebv}{\ifmmode E_{\rm B-V} \else $E_{\rm B-V}$\fi}
\newcommand{\av}{\ifmmode A_{\rm V} \else $A_{\rm V}$\fi}

\def\msun{\ifmmode M_{\odot} \else M$_{\odot}$\fi}
\def\msunyr{\ifmmode M_{\odot} {\rm yr}^{-1} \else M$_{\odot}$ yr$^{-1}$\fi}
\def\zsun{\ifmmode Z_{\odot} \else Z$_{\odot}$\fi}

\def\lsun{\ifmmode L_{\odot} \else L$_{\odot}$\fi}

\def\mup{\ifmmode M_{\rm up} \else M$_{\rm up}$\fi}
\def\mlow{\ifmmode M_{\rm low} \else M$_{\rm low}$\fi}

\def\aap{A\&A}

\def\apjl{ApJL}
\def\apjs{ApJS}

\newcommand{\oh}{\ifmmode 12 + \log({\rm O/H}) \else$12 + \log({\rm
O/H})$\fi}

\newcommand{\oii}{[O~{\sc ii}]}
\newcommand{\oiii}{[O~{\sc iii}]}

\def\Oii{[O~{\sc ii}] $\lambda$3727}
\def\Oiii{[O~{\sc iii}] $\lambda\lambda$4959,5007}

\def\flyf{\ifmmode f_{\rm Lyf} \else $f_{\rm Lyf}$\fi}
\def\pz{\ifmmode P(z) \else $P(z)$\fi}
\def\ki2{\ifmmode \chi^2 \else $\chi^2$\fi}
\def\zphot{\ifmmode z_{\rm phot} \else $z_{\rm phot}$\fi}

\newcommand{\xphot}{\ifmmode x_\gamma \else $v_\gamma$\fi}
\newcommand{\xobs}{\ifmmode x_{\rm obs} \else $x_{\rm obs}$\fi}
\newcommand{\xcmf}{\ifmmode x_{\rm CMF} \else $x_{\rm CMF}$\fi}
\newcommand{\vexp}{\ifmmode V_{\rm exp} \else $V_{\rm exp}$\fi}
\newcommand{\vmax}{\ifmmode V_{\rm max} \else $V_{\rm max}$\fi}
\newcommand{\nh}{\ifmmode N_{\rm HI} \else $N_{\rm HI}$\fi}
\newcommand{\dv}{\ifmmode \Delta v({\rm em-abs}) \else $\Delta v({\rm em}-{\rm abs})$\fi}

\def\fesc{\ifmmode f_{\rm esc} \else $f_{\rm esc}$\fi}

\def\frellya{\ifmmode f^{\rm rel}_{\rm{Ly}\alpha} \else $f^{\rm rel}_{\rm{Ly}\alpha}$\fi}

\def\hii{H{\sc ii}}

\newcommand{\mstar}{\ifmmode M_\star \else $M_\star$\fi}
\newcommand{\muv}{\ifmmode M_{1500} \else $M_{1500}$\fi}
\newcommand{\auv}{\ifmmode A_{\rm UV} \else $A_{\rm UV}$\fi}
\newcommand{\luv}{\ifmmode L_{\rm UV} \else $L_{\rm UV}$\fi}
\newcommand{\lir}{\ifmmode L_{\rm IR} \else $L_{\rm IR}$\fi}
\newcommand{\lbol}{\ifmmode L_{\rm bol} \else $L_{\rm bol}$\fi}
\newcommand{\liruv}{\ifmmode L_{\rm IR+UV} \else $L_{\rm IR+UV}$\fi}
\newcommand{\liroveruv}{\ifmmode L_{\rm IR}/L_{\rm UV} \else $L_{\rm IR}/L_{\rm UV}$\fi}
\newcommand{\nlyc}{\ifmmode N_{\rm Lyc} \else $N_{\rm Lyc} $\fi}
\newcommand{\rholyc}{\ifmmode \rho_{\rm Lyc} \else $\rho_{\rm Lyc} $\fi}
\newcommand{\chion}{\ifmmode \xi_{\rm ion} \else $\xi_{\rm ion}$\fi}
\newcommand{\chioncorr}{\ifmmode \xi_{\rm ion}^0 \else $\xi_{\rm ion}^0$\fi}

\begin{document}
    \title{The ionizing photon production efficiency of compact $z \sim 0.3$ Lyman continuum leakers and comparison with
    high redshift galaxies}
  \subtitle{}
  \author{D. Schaerer\inst{1,2}, 
Y. I. Izotov$^{3}$,
A. Verhamme$^{1}$,
I. Orlitov\'a$^{4}$,
T. X. Thuan$^{5}$, 
G. Worseck$^{6}$, 
N. G. Guseva$^{3}$
  }
  \institute{
Observatoire de Gen\`eve, Universit\'e de Gen\`eve, 51 Ch. des Maillettes, 1290 Versoix, Switzerland
         \and
CNRS, IRAP, 14 Avenue E. Belin, 31400 Toulouse, France
	\and
Main Astronomical Observatory, Ukrainian National Academy of Sciences,
27 Zabolotnoho str., Kyiv 03680, Ukraine
	\and
Astronomical Institute, Czech Academy of Sciences, Bo\v cn{\'\i} II 1401, 141 00, Prague, Czech Republic
	\and
Astronomy Department, University of Virginia, P.O. Box 400325, 
Charlottesville, VA 22904-4325, USA
	\and
Max-Planck-Institut f\"ur Astronomie, K\"onigstuhl 17, 69117 Heidelberg, Germany		
	 }

\authorrunning{D.\ Schaerer et al.}
\titlerunning{Compact $z \sim 0.3$ Lyman continuum leakers}

\date{Accepted for publication in A\&A Letters}


\abstract{We have recently discovered five Lyman continuum leaking galaxies at $z \sim 0.3$, selected for their
compactness, intense star-formation, and high \oiii $\lambda$5007/\Oii\ ratio  \citep{Izotov2016Eight-per-cent-,Izotov2016}.
Here we derive their  ionizing photon production efficiency, \chion, 
a fundamental quantity for inferring the number of photons available 
to reionize the Universe, for the first time for galaxies with confirmed strong Lyman continuum escape ($\fesc \sim 6-13$ \%).
We find an ionizing photon production per unit UV luminosity, \chion, which is a factor 2--6 times higher than the canonical 
value when reported to their observed UV luminosity. After correction for extinction this value is close to the canonical value.
The properties of our five Lyman continuum leakers are found to be very similar to those of the confirmed
$z=3.218$ leaker {\em Ion2} from \cite{de-Barros2016An-extreme-O-II} and very similar to those of typical star-forming galaxies 
at $z \protect\ga 6$. Our results suggest that UV bright galaxies at high-$z$ such as Lyman break galaxies
can be Lyman continuum leakers and that their contribution to cosmic reionization may be underestimated.
}


 \maketitle

\section{Introduction}
\label{s_intro}
In the quest for identifying the main sources of cosmic reionization and understanding 
this early epoch of the Universe, three important factors need to be quantified.
First, sources emitting Lyman continuum (LyC) photons into the inter-galactic medium (IGM)
must be identified and their emission
quantified.
Second, an average escape fraction of ionizing photons must be estimated or assumed.
And third, the total ionizing photon production of galaxies (or other sources)
needs to be related to a statistical quantity 
such as a 
luminosity function to compute the total amount of ionizing photons emitted and escaping from such a population.

Because the galaxy UV luminosity function at high-$z$ is fairly well determined
\cite[e.g.][]{Bouwens2015UV-Luminosity-F,Finkelstein2015The-Evolution-o} 
it is convenient to write the rate of ionizing photons escaping from galaxies as 
\begin{equation}
\fesc \times\ \nlyc = \fesc\ \xi_{\rm ion} \, L_\nu, 
\label{eq_nlyc}
\end{equation}
where \nlyc\  is the Lyman continuum photon production rate, \fesc\ the LyC escape fraction,
$L_\nu$ the monochromatic UV luminosity, 
and therefore 
$\chion=\nlyc/L_\nu$ 
the ionizing photon production per unit UV luminosity, i.e. ``efficiency''.
Knowing \fesc\ and $\xi_{\rm ion}$ one can thus simply compute the total photon rate at
which a given galaxy population ionizes the IGM \cite[e.g.][]{Robertson2013New-Constraints}.

The production efficiency  $\xi_{\rm ion}$ of a given stellar population is a simple
prediction from synthesis models, from which canonical values of  $\log(\chion) \approx 25.2-25.3$ erg$^{-1}$ Hz
are adopted for high-$z$ studies \citep[e.g.][]{Robertson2013New-Constraints}, 
corresponding to constant star-formation and slightly sub-solar metallicity.
Higher values may be obtained by stellar population models with young ages, non-constant star formation histories, 
lower metallicities, or when binary stars are included 
\citep[see e.g.][]{schaerer2003,Robertson2013New-Constraints,Wilkins2016The-Lyman-conti}.
Observationally \chion\ has recently been estimated by \cite{Bouwens2015Using-the-Infer} for a sample 
of high-$z$ Lyman break galaxies (LBGs) combining indirect measurements of \ha\ from photometry with the observed UV luminosity,
and in another study for a lensed $z=7$ galaxy \citep{Stark2015Spectroscopic-d}, finding values of \chion\
compatible with canonical values or somewhat higher.
However, the ionizing photon production of galaxies known to be 
LyC leakers (i.e.\ with $\fesc>0$) has not been measured so far.
 
Selecting star-forming galaxies for their compactness and high emission line ratio \oiii $\lambda$5007/\Oii=O$_{32}>5$,
 \cite[][hereafter I16a,I16b]{Izotov2016Eight-per-cent-,Izotov2016}
have recently found five $z \sim 0.3$ sources from a sample of five showing a clear detection in the 
LyC with corresponding absolute escape fractions $\fesc \sim 6-13$ \%. 
This breakthrough in the identification of LyC leakers at low-$z$ provides us now
for the first time with the opportunity to determine their ionizing photon production and other properties,
and to examine how representative these sources could be for galaxies at high redshift, 
close to and within the epoch or reionization.
In the present Letter we report the results from this analysis and comparison.

The ionizing properties of our sources are discussed in Sect.\ \ref{s_ion}.
In Sect.\ \ref{s_comp} we show that the main observed and derived properties 
of our $z \sim 0.3$ sources are very similar to those of ``typical" galaxies
at high-$z$.
Our main results are summarized in Sect.\ \ref{s_conclude}.
We adopt a Lambda-CDM cosmological model with $H_{0}$=70 km s$^{-1}$ Mpc$^{-1}$, 
$\Omega_{m}$=0.3 and $\Omega_{\Lambda}$=0.7. Magnitudes are given in the AB system. 

\section{UV and ionizing properties of $z \sim 0.3$ leakers}
\label{s_ion}
We use the GALEX and SDSS photometry as well as emission line measurements of the five 
Lyman continuum leakers reported in I16ab to determine their ionizing photon production efficiency and
other UV properties.
Since the luminosity in the optical hydrogen recombination lines is proportional to the number of 
LyC photons
absorbed in the galaxy, we determine \nlyc\ for our sources from 
\begin{equation}
\nlyc ({\rm s^{-1}})= 2.1 \times 10^{12}  (1- \fesc)^{-1} L(\hb) \,\,({\rm erg \, s^{-1}})
\label{eq_nlyc}
\end{equation}
where $L(\hb)$ is the (extinction-corrected) \hb\ luminosity from I16ab,
and the numerical coefficient translates the recombination line intensity for typical conditions
in \hii\ regions \citep{Storey95}.
Cast in terms of the absolute UV magnitude, one has 
$\log(\xi_{\rm ion}) = \log(N_{\rm Lyc}) + 0.4 \times M_{\rm  UV} - 20.64$.
The absolute UV magnitude $M_{1500}$, uncorrected for extinction, was determined from the best-fit SED to the broad-band photometry
of our sources using the fitting tool described below.
For comparison
with high-$z$ galaxy observations (cf.\ below) we also use the best-fit SED to determine the UV slope
$\beta_{1500}$\footnote{$\beta_{1500}$ is defined as the slope of the spectrum $F_\lambda \propto \lambda^\beta$ between 1300 and 1800 \AA.
$\beta_{2000}$, measured over 1800--2200 \AA, is typically bluer by $\sim 0.2-0.4$ for our sources.}.
Other data are taken from I16ab.
The most important derived quantities are summarized in Table \ref{table1}.
The main uncertainty on \chion\ comes from the aperture correction for the H$\beta$ luminosity (I16ab), 
which we estimate is $< 30-40$ \%.  We estimate the uncertainty in the extinction correction of the UV flux to 
be $\sim 30$\%. We therefore adopt a typical error of $\pm 0.1$ (0.15) dex for \chion\ (\chioncorr).

\begin{figure}[tb]
{\centering
\includegraphics[width=9cm]{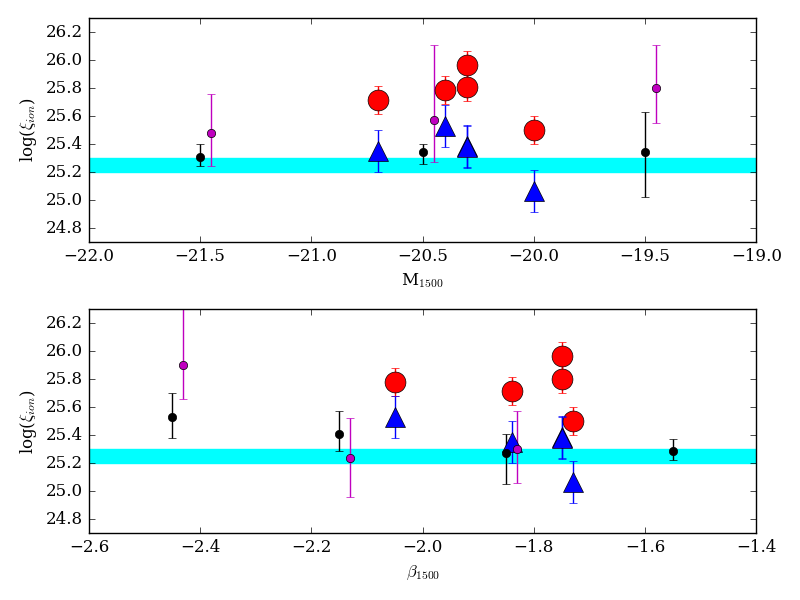}
\caption{Ionizing photon production per unit UV luminosity, \chion, as a function of the absolute UV magnitude (top panel)
and the UV slope (bottom) of the five Lyman continuum leakers of I16ab. Large red symbols show \chion, large blue symbols
\chioncorr\ after correction for  UV attenuation (two blue triangles are indistinguishable).
The cyan band illustrates canonical values for the intrinsic \chioncorr.
Recent determination of \chioncorr\ for LBGs at $z=3.8-5$ and $z=5.1-5.4$ from \cite{Bouwens2015Using-the-Infer}
are shown by small black and magenta symbols with errorbars, respectively.}
}
\label{fig_chi}
\end{figure}

In Fig.\ 1 we show the ionizing photon production efficiency (i.e.\ per UV luminosity) of our Lyman
continuum leakers as a function of UV magnitude, and compare those with the 
canonical value $\log(\chion) \approx 25.2-25.3$ erg$^{-1}$ Hz (cf.\ above),
and the recent estimate from observations of high redshift galaxies \cite[][hereafter B15]{Bouwens2015Using-the-Infer}.
Normalized to the {\em observed} UV luminosity the ionizing photon production efficiency of our
sources is found to be $\log(\chion) \approx 25.5-26$ erg$^{-1}$ Hz, i.e.\ a 
factor 2--6 times higher than the canonical value,
which is generally applied to translate the observed UV luminosity density to a global ionizing photon production 
rate \citep[e.g.][]{Robertson2013New-Constraints}. This implies that
the contribution of relatively bright galaxies, say at $\sim (0.4-1) L^\star_{\rm UV}$ for $z \sim 6-8$
\citep[cf.][]{Bouwens2015UV-Luminosity-F,Finkelstein2015The-Evolution-o},
to the cosmic ionizing photon production could be larger than commonly thought.

The UV flux of our leaking galaxies is attenuated by a factor 1.8 -- 3.8 with a median
of 2.6 ($\auv \approx 1$; cf.\ Table \ref{table1}).
After correction for dust attenuation the resulting {\em intrinsic} ionizing photon production efficiency
\chioncorr, also listed in the Table, is $\log(\chion) \approx 25.1-25.5$ (erg Hz$^{-1}$)$^{-1}$,
close to the canonical value.

The behavior of the observed and dust-corrected values of \chion\ and \chioncorr\ respectively,
now as a function of the observed UV slope, is shown in the bottom panel
of Fig.\ 1. 
Our sources have UV slopes of the order of $\beta_{1500} \sim -1.7$ to $-2$.
Broadly speaking our results are comparable to those derived by B15, who find values
of \chioncorr\ compatible with the canonical one for the bulk of their sources, and a possible
increase for the bluest sources ($\beta < -2.3$).
However, an important point to keep in mind is that the determinations of 
\chioncorr\ by B15 rely on the use of the UV slope to estimate the UV attenuation\footnote{For the SMC law 
their relation is $\auv = 1.1 ( \beta - \beta_0) = 1.1 (\beta + 2.23)$, where $\beta_0=-2.23$ is the intrinsic UV slope 
corresponding to solar metallicity and constant SFR with age $>100$ Myr.}.
For example, for sources with $\beta=-2$ and an intrinsic slope $\beta_0=-2.23$ this implies 
$\auv = 0.25$ for the SMC law, whereas our sources show a median $\auv \approx 1$ for  the same 
UV slope, a factor $\sim 2$ higher than the UV attenuation applied by B15.
In reality the UV attenuation of the high-$z$ galaxies analyzed by B15 could be 
underestimated since their true UV slope is expected to be bluer than $\beta_0=-2.23$,
as already stressed by \cite{dBSS14} and \cite{Castellano2014Constraints-on-}.
If correct, this would imply the same factor 2 downward revision of the value of \chioncorr\ of B15.

Independently of dust corrections, the ionizing emissivity needed to match cosmic reionization
is estimated to correspond to $\log(\fesc \chion)=24.5$ (24.9) erg$^{-1}$ Hz if the UV luminosity function extends down to
$\muv = -13$ ($-17$) \citep[cf.][]{Robertson2013New-Constraints,Bouwens2015Reionization-Af}.
Our sources show $\log(\fesc \chion)=24.24-24.83$ erg$^{-1}$ Hz with a median of 24.67,
a factor 1.5 larger than the above value $\log(\fesc \chion)=24.5$ erg$^{-1}$ Hz.
If the escape fraction of our sources was higher, $\fesc=0.2$ as assumed in these studies, they
would emit $\log(\fesc \chion)=24.8-25.7$ erg$^{-1}$ Hz, with a median of 25.1.
We now compare other observed and derived physical properties of our sources to those
of high redshift star-forming galaxies.

\begin{table}[htb]
{\small
\caption{Observed and derived UV and ionizing properties of our sample. Typical uncertainties on the UV slope 
$\beta_{1500}$ are $\pm 0.2$ and $\pm 0.1$ (0.15) dex on \chion\ (\chioncorr).}
\begin{center}
\begin{tabular}{lrrrrrrrr}
ID$^a$ & $z^b$ & $M_{1500}$ & $\beta_{1500}$ & $A_{UV}^b$ & \fesc$^b$ &  \chion\ & \chioncorr \\
\hline \\ 
          9 & 0.3013   & -20.3   & -1.75    & 1.44     & 0.072     & 25.96    & 25.39    \\ 
         11 & 0.3419   & -20.7   & -1.84    & 0.91     & 0.132    & 25.71    & 25.35    \\ 
         13 & 0.3181   & -20.0   & -1.73    & 1.09     & 0.056    & 25.50    & 25.07    \\ 
         14 & 0.2937   & -20.3   & -1.75    & 1.06     & 0.074    & 25.80    & 25.38    \\ 
         15 & 0.3557   & -20.4   & -2.05    & 0.64     & 0.058    & 25.78    & 25.53    \\ 
\hline
\multicolumn{9}{l}{$^a$The source IDs stand for 9=J0925+1403, 11=J1152+3400,}\\
\multicolumn{9}{l}{\hspace{1cm}13=J1333+6246, 14=J1442-0209, 15=J1503+3644}\\
\multicolumn{9}{l}{$^b$ Values taken from I16ab}
\vspace*{-0.5cm}
\end{tabular}
\end{center}
\label{table1}
}
\end{table}

\section{Comparison with high redshift galaxies}
\label{s_comp}
\subsection{Comparison with the $z=3.218$ LyC leaking galaxy {\em Ion2}}
In many respects, the properties of the compact $z \sim 0.3$ leaking sources are very comparable
to those of the $z=3.218$ Lyman continuum leaker {\em Ion2} found by 
\cite{Vanzella2015Peering-through} and \cite{de-Barros2016An-extreme-O-II}.
First {\em Ion2} is also bright in the UV, $M_{\rm UV}=-21$ which is $\approx M^\star_{\rm UV}(z=3)$.
Its low stellar mass, $\la 1.6 \times 10^9$ \msun, is comparable to $\mstar = (0.2 - 4) \times 10^9$ \msun\
(a median of $1 \times 10^9$ \msun) of our five sources (I16ab).
The metallicity of  {\em Ion2}, determined from rest-frame UV and optical emission lines is $\sim 1/6$ solar, compared to $\sim (0.1-0.2)$ solar.

The non-detection of \Oii\ translates to a 2-$\sigma$ lower limit of a high ratio O$_{32}>10$
for {\em Ion2}, even higher than for our objects. Furthermore, {\em Ion2} is also
a compact source with a size $\sim 300 \pm 70$ pc  \citep[cf.][]{de-Barros2016An-extreme-O-II}.
Interestingly the two latter properties are found {\em a posteriori}, since the source was selected from 
a peculiar color selection. 
Different selection criteria finding LyC leakers may thus pick up sources with similar properties.

Finally, {\em Ion 2} shows strong rest-frame optical emission lines, e.g.\ EW(5007)$=1103 \pm 60$ \AA\
or even larger by a factor $\ga 2$ if corrected for a large escape fraction of ionizing photons, 
compared to EW(5007)$=900-1260$ \AA\ of the $z\sim0.3$ leakers (I16ab).
Such high equivalent widths seem fairly typical for star-forming galaxies at $z \ga 6$, as we will now see
(cf.\ Fig.\ 3).

\subsection{Comparison with typical high-$z$ galaxies}
As clear from Table \ref{table1}, our five Lyman continuum leakers show, at the same absolute UV magnitude, 
a UV slope in good agreement with the average slope observed in Lyman break galaxies at $z \sim 4-7$ 
\cite[cf.\ e.g.][]{Dunlop2012A-critical-anal,Bouwens2013}.
Our $z \sim 0.3$ sources are also in line with the average relation between stellar mass and UV magnitude
derived at high redshifts, which indicates a stellar mass $\mstar \sim 10^9$ \msun\ for $\muv \sim -20$
\citep[cf.][]{Duncan2014The-mass-evolut,dBSS14,Grazian2015The-galaxy-stel}.

Figure 2 
shows for illustration the SED fit of the first of our LyC leakers,
J0925+1403 from I16a, to the broad-band photometry from the SDSS and the two GALEX bands. 
The fits are compatible with the more detailed SED fits to the observed COS and SDSS spectra 
discussed in I16ab.
The fit has been obtained with a version of the {\em Hyperz} code including nebular emission,
described in \cite{schaerer&debarros2009,schaerer&debarros2010} and which has extensively been used to fit large samples
of high-$z$ LBGs \citep[cf.][]{dBSS14}. Since the attenuation law and the metallicity are constrained/measured
(I16a), we have used the SMC law and a metallicity=1/5 solar, the closest value available 
for the \cite{BC03} models. Clearly the SDSS photometry is  dominated by strong emission lines in bands
at $\lambda \ga 5500$ \AA, and the SED is well fitted with the average emission line ratios taken from
\cite{Anders03} (here for 1/5 solar metallicity) and adopted in our models. This demonstrates that the SED of extreme,
rare objects of the nearby Universe with very strong emission lines can also be well reproduced with typical 
line ratios of low-$z$ galaxies.

\begin{figure}[tb]
{\centering
\vspace*{-1.4cm}
\includegraphics[width=8cm]{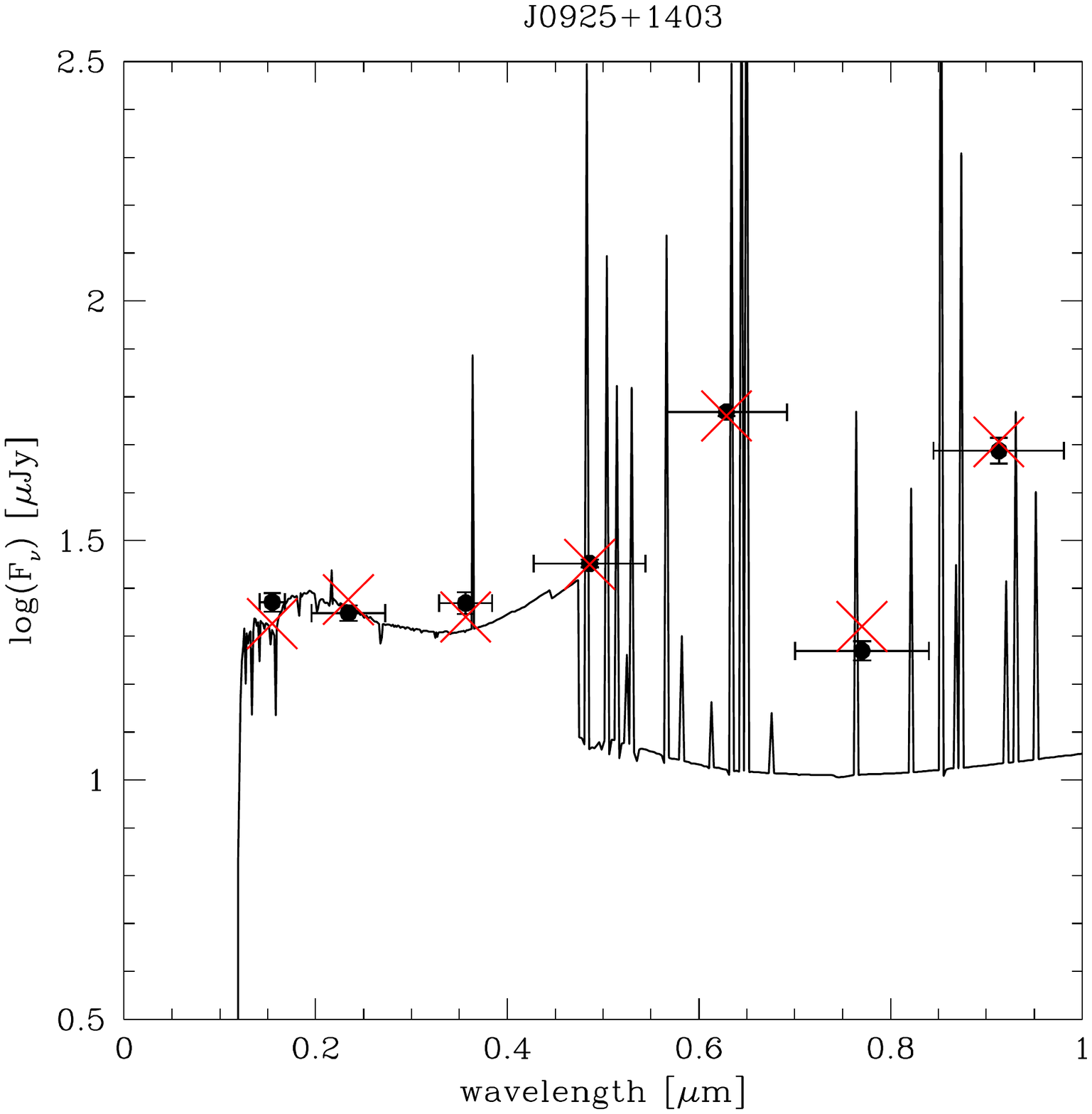}
\vspace*{-2.2cm}
\caption{Observed  broad band photometry and best-fit SED (black curve) of the compact Lyman continuum leaker 
J09 from I16a. Red crosses indicate  the synthetic flux in the corresponding filters, showing that most of the optical
bands are dominated by nebular emission.}
}
\label{fig_sed}
\end{figure}

A salient feature of LBGs at high redshift, which has clearly been established during recent years,
is the presence of strong optical emission lines, whose signature is detected in broad band photometry
and whose (average) strength increases strongly with redshift
\citep[e.g.][]{shimetal11,Labbe2013The-Spectral-En,dBSS14}.
We now compare the line strengths (equivalent widths) of our $z \sim 0.3$ 
LyC leakers with these observations, summarised in 
Figure 3 for \ha, \oiii $\lambda$5007, and \oii\ with fits derived by \cite{Khostovan2016The-Nature-of-H}
for the range of stellar mass  $9.5 < \log(\mstar/\msun)<10$, where these quantities can be
derived over a wide redshift domain.
Overplotted is the range of EWs measured in our five $z \sim 0.3$ LyC leakers
reported in I16ab. With rest-frame EW(\ha)$\sim 560-1060$ \AA\ and EW(\oiii $\lambda$5007) $\sim 900-1260$ \AA\ our sources
are comparable to typical star-forming galaxies at $z \ga 6$\footnote{For galaxies with a median mass $\sim 10^9$ \msun\
as our sources, such EWs may be typical at somewhat lower redshift already, since EWs increase on average with decreasing
stellar mass \cite[cf.][]{Khostovan2016The-Nature-of-H}.}. This shows that the nebular properties of 
extreme and rare objects of the low redshift Universe, selected by compactness and high O$_{32}$ ratios,
appear to be very similar of those of average star-forming galaxies at high-$z$.
By analogy with our leakers, this also suggests that Lyman continuum leaking may be frequent in 
high redshift galaxies.

From Fig.\ 3 we note also that the observed EW(\Oii) $\sim 60-130$ \AA\ of our
sources agrees also well with the behavior of \Oii\ extrapolated to high redshift by \cite{Khostovan2016The-Nature-of-H}
for the same redshift ($z \sim 6-8$) as indicated by the other emission lines.
If true, this would indicate that the average O$_{32}$ ratio continues to increase beyond $z \ga 4$,
continuing the trend already observed from $z \sim 0$ to 3, as already shown by \cite{Khostovan2016The-Nature-of-H}.
If the current small samples are representative, high O$_{32}$ ratios (say O$_{32}>4$) imply a LyC escape 
fraction $\fesc > 5$\% with a possible trend of \fesc\ increasing with O$_{32}$ (I16b).
This would imply that the average star-forming galaxy at $z>4$ would also be a leaker with $\fesc > 5$ \%, 
since the average O$_{32}$ ratio exceeds 4 above this redshift \citep[cf.][]{Khostovan2016The-Nature-of-H}.

\begin{figure}[tb]
\hspace*{-0.5cm}
{\centering
\includegraphics[width=10cm]{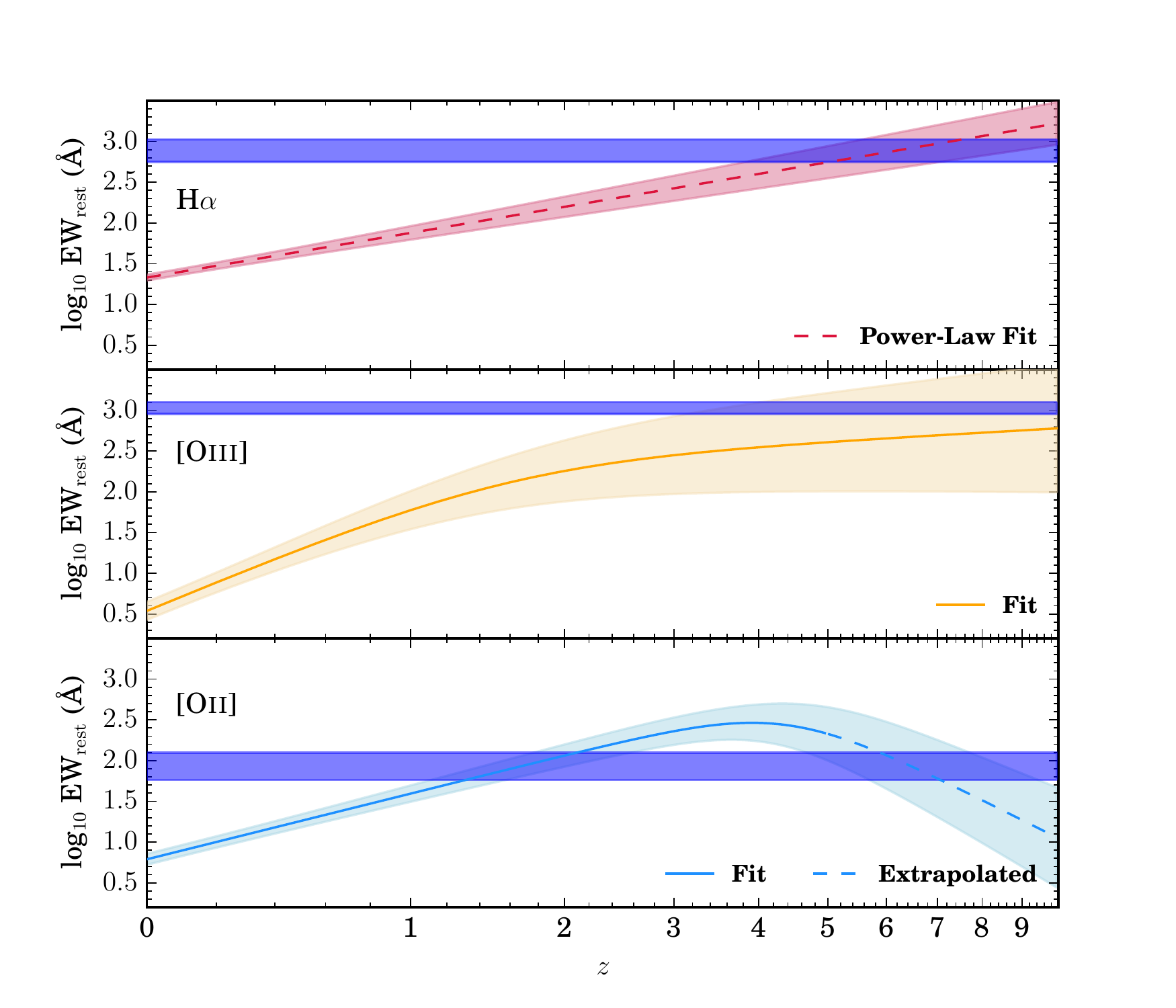}
\caption{Top, middle and bottom panels show the redshift evolution of the average (rest-frame) equivalent widths of
\ha, \oiii $\lambda$5007, and \oii\ $\lambda$3727 respectively for galaxies with stellar masses $9.5 < \log(\mstar/\msun)<10$ 
as fitted by \cite{Khostovan2016The-Nature-of-H},
and the range of the EWs observed in our five $z \sim 0.3$ Lyman continuum leakers (blue horizontal bands).
The leakers from our study show line equivalent widths typical for star-forming galaxies at $z \protect\ga 6$.}
}
\label{fig_ew}
\end{figure}

\subsection{Discussion}
All the above mentioned properties show that the five $z \sim 0.3$ galaxies recently identified
as Lyman continuum leakers by I16ab are very similar to both
the arguably most reliable high-$z$ leaker, the $z=3.218$ galaxy found by
\cite{Vanzella2015Peering-through} and \cite{de-Barros2016An-extreme-O-II}, and to
typical star-forming galaxies at $z \ga 6$.

In terms of their very high equivalent widths of \Oiii\ and \ha\ our $z\sim 0.3$ sources are very 
rare for low-$z$ sources. In fact by this measure they correspond to the $<10^{-3}$ tail of the high EW distribution
of SDSS DR12 galaxies. On the other hand these high equivalent widths appear to be common,
possibly even typical, for $z>6$ LBGs, as shown in Fig.\ 3. 
This analogy with our LyC leakers suggests that the typical LBG at high redshift may
also be leaking Lyman continuum radiation.

The recent work of \cite{Sharma2016The-brighter-ga} provides indirectly further support for this
hypothesis. Indeed making simple, but plausible assumptions about local  LyC escape,
these authors predict the escape fraction for galaxies from simulations, finding an increasing
\fesc\ with redshift and significant escape from most high-$z$ galaxies. They trace these trends 
back to the mean surface density of star formation, which is found to be very high in most of their
simulated galaxies at high redshift.
Interestingly all five $z \sim 0.3$ LyC leakers from I16ab also show a very high 
surface density of star formation, $\Sigma_{\rm SFR} \sim 2-50$ \msunyr kpc$^{-2}$, comparable
to observations of high-$z$ galaxies and to the predictions of \cite{Sharma2016The-brighter-ga}.
The success of the selection method of I16ab in finding LyC leakers at high O$_{32}$ ratios may thus be related 
to compactness, which -- together with very strong emission lines indicating a high specific SFR --
implies a high surface density of star formation. 
\cite{Heck01,Heckman2011Extreme-Feedbac} and \cite{Borthakur2014A-local-clue-to} have already suggested that such 
strong star formation  results in strong outflow, which clear channels in the ISM allowing thus
the escape of Lyman continuum photons.

\section{Conclusion}
\label{s_conclude}

We have analyzed the properties of five low redshift Lyman continuum leaking galaxies observed with the COS spectrograph onboard HST
and reported recently by \cite{Izotov2016Eight-per-cent-,Izotov2016}.
The $z \sim 0.3$ sources have been selected for compactness and for showing a high emission line ratio O$_{\rm 32}$, which has
previously been suggested as a possible diagnostic for Lyman continuum escape \citep{Jaskot2013The-Origin-and-,Nakajima2014Ionization-stat}.

We have determined the ionizing photon flux production of these galaxies which are metal-poor ($\sim 1/6$ solar),  
dominated by young stellar populations ($< 10$ Myr; cf.\ I16ab), and which are relatively UV bright 
($M_{1500} \sim -20$ to $-20.8$, cf.\ Table \ref{table1}).
Finally we have compared the observed and derived physical properties of these rare, extreme objects from the 
nearby Universe to those of high redshift galaxies.
Our main results can be summarized as follows:
\begin{itemize}
\item The ionizing photon production efficiency per observed UV luminosity, 
\chion, of the leakers is $\log(\chion) \approx 25.6-26$ erg$^{-1}$ Hz, 
i.e.\ a factor 2--6 times higher than the canonical value \citep[cf.][]{Robertson2013New-Constraints,Wilkins2016The-Lyman-conti}.
\item Although our sources show a low extinction in the optical ($A_V \sim 0.15-0.4$) their UV attenuation is  $A_{UV} \sim 0.6-1.4$,
which implies that the intrinsic, extinction-corrected ionizing photon flux production efficiency is $\log(\chioncorr) \approx 25.0-25.6$ erg$^{-1}$ Hz,
close to the canonical value.
\item The five $z\sim 0.3$ sources of I16ab share many properties with the best established high-$z$ leaking galaxy
{\em Ion2} \citep{de-Barros2016An-extreme-O-II}: absolute UV magnitude, stellar mass, metallicity, line equivalent widths,
high O$_{\rm 32}$ ratio and others are very similar. 
\item The high rest-frame equivalent widths of \ha\ and  \oiii $\lambda$5007 
of the $z \sim 0.3$ leakers 
are very similar to those inferred for typical star-forming galaxies at $z \ga 6$ from broad-band photometry.
This shows that the rare, extreme galaxies selected by I16ab from the Sloan survey could very well be
fairly representative of average galaxies in the early Universe. 
\item Our results also suggest that UV bright galaxies at high-$z$ such as Lyman break galaxies can be Lyman continuum leakers and 
that their contribution to cosmic reionization, based on canonical assumptions for \chion, is  likely underestimated.
\end{itemize}

\begin{acknowledgements}
We thank various colleagues, including Eros Vanzella, Andrea Grazian, and Tom Theuns for stimulating discussions,
Ali Khostovan for kindly sharing python scripts, and Yves Revaz for help with python.

\end{acknowledgements}
\bibliographystyle{aa}
\bibliography{merge_misc_highz_literature}

\end{document}